\documentclass[aps,reprint,superscriptaddress,showpacs,prl]{revtex4-1} 

\usepackage{amsmath}
\usepackage{amssymb}
\usepackage{graphicx}
\usepackage{bm}       
\usepackage{hyperref} 
\hypersetup{
     colorlinks=true,       
     linkcolor=blue,        
     citecolor=blue,        
     filecolor=blue,        
     urlcolor=cyan          
}

\newcommand{\bra}[1]{\left< #1 \right|} 
\newcommand{\ket}[1]{\left| #1 \right>} 
\newcommand{\HS}{\ensuremath{H_S}}
\newcommand{\HR}{\ensuremath{H_\mathrm{R}}}
\newcommand{\up}{\ensuremath{\uparrow}}
\newcommand{\down}{\ensuremath{\downarrow}}
\newcommand{\av}[1]{\ensuremath{\left\langle #1 \right\rangle}}

\renewcommand{\vec}[1]{\bm{#1}}

\begin{document}
\title{Backscattering in helical edge states from a magnetic impurity
  and Rashba disorder}

\author{Lukas Kimme} \affiliation{Institut f\"ur Theoretische
Physik, Universit\"at Leipzig, D-04103 Leipzig, Germany}

\author{Bernd Rosenow} \affiliation{Institut f\"ur Theoretische
Physik, Universit\"at Leipzig, D-04103 Leipzig, Germany}

\author{Arne Brataas} \affiliation{Department of Physics, Norwegian
  University of Science and Technology, NO-7491 Trondheim, Norway}

\date{\today}

\begin{abstract}
 
  Transport by helical edge states of a quantum spin Hall insulator is
  experimentally characterized by a weakly temperature-dependent mean
  free path of a few microns and by reproducible conductance
  oscillations, challenging proposed theoretical explanations. We
  consider a model where edge electrons experience spatially random
  Rashba spin-orbit coupling and couple to a magnetic impurity with
  spin $S \geq 1/2$.  In a finite bias steady state, we find for
  $S>1/2$ an impurity induced resistance with a temperature dependence
  in agreement with experiments.  Since backscattering is elastic,
  interference between different scatterers possibly explains
  conductance fluctuations.
  
\end{abstract}


\maketitle

\emph{Introduction:} During the last decade, the quantum spin Hall
effect (QSHE) \cite{Kane2005a, Kane2005b, Bernevig2006a} has become an
important example of a topologically ordered state with time-reversal
invariance.  One of its key features is the existence of helical edge
states \cite{Wu2006} with a quantized conductance of $e^2/h$ per edge,
as edge electrons are protected from elastic single particle
backscattering by time-reversal symmetry \cite{Kane2005a, Wu2006,
  Xu2006}.  Soon after the theoretical prediction
\cite{Bernevig2006b}, the QSHE was realized in HgTe/CdTe quantum wells
\cite{Koenig2007}, and the quantized conductance \cite{Koenig2007} as
well as the demonstration of nonlocal transport \cite{Roth2009} were
crucial signatures for this first-time experimental observation.
However, already in this first as well as in subsequent experiments
\cite{Koenig2007, Roth2009, Nowack2013, Grabecki2013, Gusev2014,
  Yacoby:private}, deviations from the quantized conductance with a
weak temperature dependence were found for edges longer than
approximately $1\,\mu$m.  Moreover, in short samples, where the
conductance is essentially quantized, small conductance fluctuations
are observed as the back-gate voltage is tuned \cite{Koenig2007,
  Koenig2008, Roth2009, Yacoby:private}.  After the prediction
\cite{Liu2008} of the QSHE in InAs/GaSb/AlSb quantum wells, the same
qualitative behavior of the conductance as in HgTe/CdTe was observed
also in these devices \cite{Knez2011, Suzuki2013, Knez2014,
  Spanton2014}.

A multitude of other mechanisms beyond elastic single particle
backscattering have been proposed as possible explanations for the
relatively short mean free path \cite{Dolcetto2015}: inelastic single
particle \cite{Schmidt2012, Budich2012, Lezmy2012, Kainaris2014} and
two-particle backscattering \cite{Kane2005a, Wu2006, Xu2006,
  Crepin2012, Geissler2014, Kainaris2014} which can be caused by
electron-electron or electron-phonon interactions, both of which are
usually considered in combination with other time-reversal invariant
perturbations; tunneling of electrons into charge puddles caused by
inhomogeneous doping, giving rise to inelastic single particle
backscattering \cite{Vayrynen2013, Vayrynen2014}; coupling of edge
electrons to a spin bath which gets dynamically polarized
\cite{Lunde2012}, thus effectively breaking time-reversal symmetry and
giving rise to elastic backscattering in conjunction with Rashba
disorder \cite{DelMaestro2013}; time-reversal symmetry breaking by an
exciton condensate \cite{Pikulin2014}; and coupling of edge electrons
to a single Kondo impurity \cite{Wu2006, Maciejko2009, Tanaka2011}, to
a lattice of Kondo impurities \cite{Maciejko2012}, to a single Kondo
impurity in the presence of homogeneous Rashba spin-orbit coupling
\cite{Eriksson2012, Eriksson2013}, or to several Kondo impurities with
random anisotropies \cite{Altshuler2013}.  Although these mechanisms
are very diverse, many of them have in common a pronounced temperature
dependence, usually some power law $T^\alpha$ with positive exponent
$\alpha$ for the resistance.  However, only a weak temperature
dependence has been observed experimentally, with the exception of a
recent study using very low excitation currents \cite{Li2015}.  In
fact, in some experiments, a slight increase of the resistance is
observed when the temperature is decreased \cite{Koenig2007,
  Gusev2014, Yacoby:private}.  With respect to the conductance
fluctuations, only charge puddles \cite{Vayrynen2013, Vayrynen2014} as
well as coherent scattering between several magnetic impurities with
large spin $S>1$ and uniaxial single-ion anisotropy
\cite{Cheianov2013} were considered as possible explanations.
Theories that build on scattering from local disorder are also
supported by recent scanning gate microscopy experiments
\cite{Koenig2013}, which identified individual scattering centers.

In this work, we consider scattering of helical edge electrons from a
magnetic impurity with spin $S\geq1/2$ in combination with a spatially
fluctuating Rashba spin-orbit coupling.  The latter originates from a
fluctuating electric field in the out-of-plane direction due to
disorder in the doping layers \cite{Sherman2003, Golub2004, Rothe2010,
  Glazov2010, Strom2010}. From a $T$-matrix calculation accounting for
the combined scattering events off these perturbations, we derive an
effective additional coupling to the impurity.  This coupling provides
a backscattering mechanism which is enhanced by an increased
polarization of the impurity with spin $S>1/2$.  The polarization of
the impurity spin is determined from the steady state solution to a
semiclassical scattering rate equation.  We consider the linear and
the nonlinear regime.  Upon entering the nonlinear regime with the
source drain voltage larger than temperature, the impurity gets
polarized and the Rashba disorder induced effective coupling leads to
an increased resistance, thus providing a possible explanation for the
experimental results.  We assume that the relevant Kondo temperature
is exponentially suppressed and well below the temperature regime
studied in our analysis.  Since the dominant contribution to
backscattering is elastic in our model, quantum interference between
different scatterers is possible, and can give rise to conductance
fluctuations as observed in Refs.~\cite{Koenig2007, Koenig2008, Roth2009,
  Yacoby:private}.

\emph{Model:} The edge states are described by 
\begin{equation}
  H_0 = \int dx\, \Psi_\alpha^\dagger(x)
  \sigma_{\alpha\beta}^z (-iv\hbar\partial_x) \Psi_\beta(x),
\end{equation}
where $\Psi_\up(x)$ annihilates a right-moving electron, $v$ is the
edge velocity, and the spin quantization axis is in the $z$ direction.
A disordered Rashba spin-orbit coupling is described by
\cite{Sherman2003, Golub2004, Strom2010, Rothe2010}
\begin{equation}
  \HR = \int dx\, \Psi_\alpha^\dagger(x)
  \sigma_{\alpha\beta}^y \{a(x),i\partial_x\} \Psi_\beta(x) \ \ ,
  \label{eq:HR}
\end{equation} 
with the correlator $\av{a(x)a(x')}_\mathrm{dis} = V_0F(x-x')$ and
$F(0)=1$ \cite{SuppMat,Glazov2010}.  Being time-reversal invariant,
\HR\ does not cause elastic backscattering \cite{Kane2005a}.  The
essential ingredient is the coupling of electrons to a local magnetic
impurity with spin $S$ via
\begin{equation}
  \HS = J_zS^zs^z + J_\perp(S^+s^-+S^-s^+) + J_\mathrm{aniso}(S^+ + S^-)s^z.
  \label{eq:HS}
\end{equation}
As usual, $S^\pm = S^x \pm iS^y$, and
$s^i = \Psi_\alpha^\dagger(0) \sigma^i_{\alpha\beta}\Psi_\beta(0)$ are
the local spin density operators of the edge electrons.  For
$J_\mathrm{aniso}=0$, \HS\ describes a Heisenberg XXZ coupling, which
has an axial rotation symmetry and was the subject of earlier studies
\cite{Wu2006, Maciejko2009, Tanaka2011, Maciejko2012, Eriksson2012,
  Eriksson2013}.  In systems with axial symmetric \HS, the
$z$ component of the total spin is conserved and the dc conductance is
exactly quantized \cite{Tanaka2011}.  A finite $J_\mathrm{aniso}$
breaks the axial rotation symmetry, thus enabling persistent
backscattering in the dc limit \cite{footnote:Janiso}.

In general, a coupling to the impurity spin could also involve terms
such as $S^+s^+ + S^-s^-$ and $S^z(s^++s^-)$ that break axial rotation
symmetry.  However, a microscopic analysis, based on an isotropic
$sp$-$d$ exchange interaction of bulk electrons with the impurity,
results only in the terms in Eq.~\eqref{eq:HS}, at least for HgTe/CdTe
quantum wells \cite{SuppMat}.  Nevertheless, as we will explain, the
combined processes of impurity and Rashba disorder scattering
effectively give rise to such additional terms.

\begin{figure}
  \includegraphics[width=.9\columnwidth]{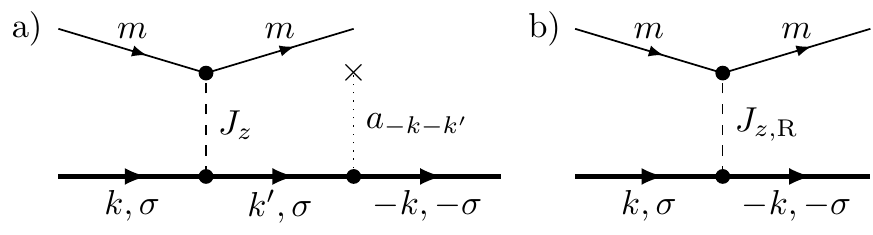}
  \caption{\label{fig:diagram} a) Diagrammatic representation of the
    second-order scattering process that gives rise to elastic
    backscattering without an impurity spin flip
    [cf. Eq.~\eqref{eq:MatrixElement}].  b) The corresponding
    first-order process from the effective coupling $\HS'$
    [cf. Eq.~\eqref{eq:HSprime}].  The dashed line denotes the
    (effective) coupling of the impurity and electron. The dotted line
    with a cross represents the Rashba potential.}
\end{figure}

\emph{Effective couplings from second-order processes:} We now
consider the combined scattering from the impurity and the Rashba
disorder by using the $T$-matrix
$T(\epsilon) = (\HS+\HR) + (\HS+\HR)G(\epsilon)T$, with
$G(\epsilon)=(\epsilon-H_0)^{-1}$.  The $T$-matrix element associated
with the first term of Eqs.~\eqref{eq:HS} and \eqref{eq:HR} is
\begin{equation}
\begin{split}
  \bra{m;-k,-\sigma}T_{z,\mathrm{R}}(\epsilon_i) \ket{m;k,\sigma} = \\
  \frac{\sigma i J_z \bra{m}S^z \ket{m}}{L\sqrt{L}}
  \sum_{k'} \left( \frac{(k-k')a_{-k-k'}}{\hbar v(k-k')} + 
    \frac{(k'+k)a_{k'-k}}{\hbar v(k'+k)} \right)
    \label{eq:MatrixElement}
\end{split}
\end{equation}
(cf. Fig.~\ref{fig:diagram}).  Here, $\ket{m;k,\sigma}$ denotes a
product state of the local moment in the $S^z$ eigenstate $\ket{m}$
and an electron with helical spin $\sigma$ and momentum $\hbar k$, $L$
the distance between the left and right reservoir,
$\epsilon_i = \sigma v\hbar k$ the energy of the initial state
$\ket{m;k,\sigma}$, and $a_q$ the Fourier components of $a(x)$.  The
energy difference between the initial and intermediate state in the
denominator compensates the matrix element from Rashba disorder in the
numerator.

With regard to scattering rates, to be discussed below, the
second-order process described in Eq.~\eqref{eq:MatrixElement} can
effectively be described as a first-order process resulting from the
additional anisotropic coupling
\begin{equation}
  \HS' = J_{z,\mathrm{R}}S^z(s^+ + s^-).
  \label{eq:HSprime}
\end{equation}
Here, $J_{z,\mathrm{R}} = 2\sqrt{\eta} J_z$ is an effective coupling
constant with $\eta = V_0 / \hbar^2 v^2$.  Analogously to the combined
process described by $T_{z,\mathrm{R}}$, there are also second-order
processes where electrons scatter from the Rashba disorder and from
the impurity by either the $J_\perp$ or the $J_\mathrm{aniso}$ term of
$\HS$ in Eq.~\eqref{eq:HS}.  Again, these effects can be captured by
considering first-order processes from the effective couplings
$J_{\perp,\mathrm{R}} (S^+ + S^-) s^z$ and
$J_{\mathrm{aniso},\mathrm{R}} (S^+s^+ + S^+s^- + \mathrm{H.c.})$,
with $J_{\perp,\mathrm{R}} = \sqrt{\eta} J_\perp$ and
$J_{\mathrm{aniso},\mathrm{R}} = 2\sqrt{\eta} J_\mathrm{aniso}$,
respectively.  $J_{\perp,\mathrm{R}}$ only renormalizes the
$J_\mathrm{aniso}$ coupling already present in \HS.  In contrast,
$J_{\mathrm{aniso},\mathrm{R}}$, besides renormalizing $J_\perp$,
opens an additional scattering channel via the $S^+s^+ + S^-s^-$
terms.

\emph{Scattering rates:} Since we are interested in the effect of the
impurity on dc transport, we pursue two aims: (i) We want to achieve a
description of the impurity spin in a driven steady state, and (ii) we
want to compute the transport scattering rates determining the
impurity induced dc resistance.  To achieve (i), we calculate the
integrated rates $\Gamma_{m'm}^{\sigma'\sigma}$, which characterize
the scattering of an electron from an initial state with helical spin
$\sigma$ into a final state with helical spin $\sigma'$, as the
impurity spin is flipped from $\ket{m}$ to $\ket{m'}$.
$\Gamma_{m'm}^{\sigma'\sigma}$ are obtained by weighting the rates for
scattering from $\ket{m;k,\sigma}$ to $\ket{m';k',\sigma'}$ with the
probability for the initial and final state to be occupied or
unoccupied, respectively, and then summing over initial and final
momenta.  To calculate the individual rates, we employ Fermi's golden
rule for $\HS^\mathrm{eff}$, including all original and effective
impurity couplings.  For temperatures and voltages much smaller than
the bulk excitation gap of the topological insulator, the weak
momentum dependence of
$\bra{m';k',\sigma'} \HS^\mathrm{eff} \ket{m;k,\sigma}$ can be
neglected.  For forward scattering with $\sigma' = \sigma$, we then
find
\begin{equation}
  \Gamma_{m'm}^{\sigma\sigma}(\beta) = \frac{L^2}{2\pi \hbar^3 v^2 \beta}
  |\bra{m'; \sigma}\HS^\mathrm{eff} \ket{m; \sigma}|^2,
  \label{eq:GammaForward}
\end{equation}
where the temperature dependence is due to the integrated occupation
factors $\int dE\, f_{\sigma}(1-f_{\sigma}) = 1/\beta$.  Here, $f_\up$
($f_\down$) denotes the Fermi distribution function describing the
occupation of right-moving (left-moving) electrons from the left
(right) reservoir.  In the case of backscattering $\sigma' = -\sigma$,
the rates involve occupation factors $f_\sigma(1 - f_{-\sigma})$,
leading to a voltage dependence
\begin{equation}
  \Gamma_{m'm}^{-\sigma\sigma}(\beta,eV) = \frac{L^2}{2\pi \hbar^3 v^2 \beta}
  |\bra{m'; -\sigma}\HS^\mathrm{eff} \ket{m; \sigma}|^2 I^\sigma(\beta eV),
  \label{eq:GammaBackward}
\end{equation}
where
\begin{equation}
  I^\sigma \equiv \beta \int dE\, f_{\sigma}(1-f_{-\sigma})
    = \sigma\beta eV \frac{e^{\sigma\beta eV}}{e^{\sigma\beta eV}-1}.
\end{equation}
At low bias voltage, $\beta eV\ll 1$, $I^\sigma \simeq 1$ so that the
forward and backscattering rates have the same temperature dependence.
On the other hand, when $\beta eV \gg 1$, the backscattering of right
movers is linearly enhanced, while the backscattering of left movers
is exponentially suppressed.  For detailed results, see
Table~\ref{tab:processes}.
\begin{table}
  \caption{
    \label{tab:processes}
    Results for the integrated scattering rates
    $\Gamma_{mm'}^{\sigma\sigma'}$.}
  \begin{ruledtabular}
    \begin{tabular}{lc}
      Process & $\Gamma \times (2\pi\hbar^3 v^2 \beta / L^2)$\\\hline
      $\ket{m;\sigma}\rightarrow\ket{m;-\sigma}$ & 
                 $J_{z,\mathrm{R}}^2 |\bra{m}S^z \ket{m}|^2 I^\sigma$ \\
      $\ket{m;\up}\rightarrow\ket{m+1;\down}$ &
                 $(J_\perp^2 + J_{\mathrm{aniso},\mathrm{R}}^2)|\bra{m+1}S^+ \ket{m}|^2 I^\up$ \\
      $\ket{m;\up}\rightarrow\ket{m-1;\down}$ &
                 $J_{\mathrm{aniso},\mathrm{R}}^2 |\bra{m-1}S^- \ket{m}|^2 I^\up$ \\
      $\ket{m;\sigma}\rightarrow\ket{m\pm1;\sigma}$ &
                 $(J_\mathrm{aniso}^2 + J_{\perp,\mathrm{R}}^2) |\bra{m\pm 1}S^\pm \ket{m}|^2$
    \end{tabular}
  \end{ruledtabular}
\end{table}

\emph{Master equation:} We describe the state of the impurity by a
density matrix $\rho$ and assume that dephasing from the coupling to
the electron bath is sufficiently strong, such that we can neglect
coherences and consider $\rho = \sum_{m} P_m\ket{m}\bra{m}$ to be
diagonal in the basis of eigenstates of $S^z$,
$S^z\ket{m} = m\ket{m}$.  We can then proceed to determine the steady
state of the impurity spin at finite temperature and under an applied
transport voltage $V$ from the master equation
\begin{equation}
  \partial_tP_{m} = \sum_{m'} (\Gamma_{mm'}P_{m'} - \Gamma_{m'm}P_m),
  \label{eq:MasterEquation}
\end{equation}
where
$\Gamma_{m'm} = \sum_{\sigma'\sigma}\Gamma_{m'm}^{\sigma'\sigma}$.  In
a steady, state $\partial_tP_{m} = 0$, and we find the recursively
defined solution $P_{m-1} = \zeta P_m$, with
$\zeta = \Gamma_{m-1,m}/\Gamma_{m,m-1}$.  $\zeta$ depends on
$\beta eV$ and the (effective) coupling constants, but not on $m$
because the $m$-dependent matrix elements of the ladder operators
$S^\pm$ cancel.  We provide an explicit expression in the supplemental
material \cite{SuppMat}.  For the model considered here, we have
$0\leq \zeta \leq 1$ and the two limiting cases have simple solutions:
$\zeta = 0$ implies $P_m = \delta_{m,S}$ and corresponds to a
maximally polarized local moment, while for $\zeta = 1$ the impurity
is completely unpolarized, i.e., $P_m = 1/(2S+1)$.  Notice that
$\beta eV=0$ implies $\zeta=1$, because, without an applied transport
voltage, there is no asymmetry between the rates for forward and
backscattering [cf. Eqs.~\eqref{eq:GammaForward} and
\eqref{eq:GammaBackward}].  The general dependence of $P_m$ on $m$
interpolates between the two limiting cases:
\begin{equation}
  P_{m} = \frac{(1-\zeta)\zeta^S}{1-\zeta^{1+2S}} \left( \frac{1}{\zeta} \right)^m.
  \label{eq:pmSteadyState}
\end{equation}

\emph{Impurity induced resistance:} The backscattering probability $R$
can be related to a scattering rate $1/\tau$ by multiplying with the
time of flight $L/v$. This compensates a dependence
$1/\tau \propto 1/L$ in all scattering rates, due to the normalization
of the plane wave with a factor $1/\sqrt{L}$.  In this way, both $R$
and the edge conductance $G =(1 - R) e^2/h $ are independent of the
system size $L$, as expected for a single scattering site.  For
$R\ll 1$, $R$ equals the impurity induced resistance normalized by
$h/e^2$.  From the effective impurity coupling we have two important
backscattering mechanisms \cite{footnote:tau2}, and hence obtain
  \begin{equation}
  R = \frac{L}{v}\left( \frac{1}{\tau_{z,\mathrm{R}}} +
      \frac{1}{\tau_\perp} \right),
    \label{eq:resistance}
  \end{equation}
where a Fermi's golden rule
calculation 
yields
\begin{subequations}
\begin{align}
    &\frac{1}{\tau_\perp} = \frac{J_\perp^2 + J_{\mathrm{aniso},\mathrm{R}}^2}{\hbar^2 vL}
              \,p_\perp \!\!\! \sum_{m=-S}^{S-1} \! |\bra{m+1}S^+\ket{m}|^2P_{m} \label{eq:tauperp}\\
    &\frac{1}{\tau_{z,\mathrm{R}}} = \frac{J_{z,\mathrm{R}}^2}{\hbar^2 vL}
                      \sum_{m=-S}^S |\bra{m}S^z\ket{m}|^2 P_{m} \label{eq:tauzR} \ \ .
\end{align}
\end{subequations}
Here, $p_\perp = 1 - \Gamma_{m,m+1}^{\up\down}/\Gamma_{m,m+1}$
accounts for the fact that the dc resistance is affected only by those
backscattering events of right-moving electrons which are \emph{not}
compensated by a subsequent backscattering of a left-moving electron
\cite{SuppMat}.  For example, in the case of vanishing Rashba disorder
and \HS\ with axial rotation symmetry, we find
$\Gamma_{m,m+1}^{\up\down} = \Gamma_{m,m+1}$, hence $p_\perp$
vanishes.

\begin{figure}
  \includegraphics[width=\columnwidth]{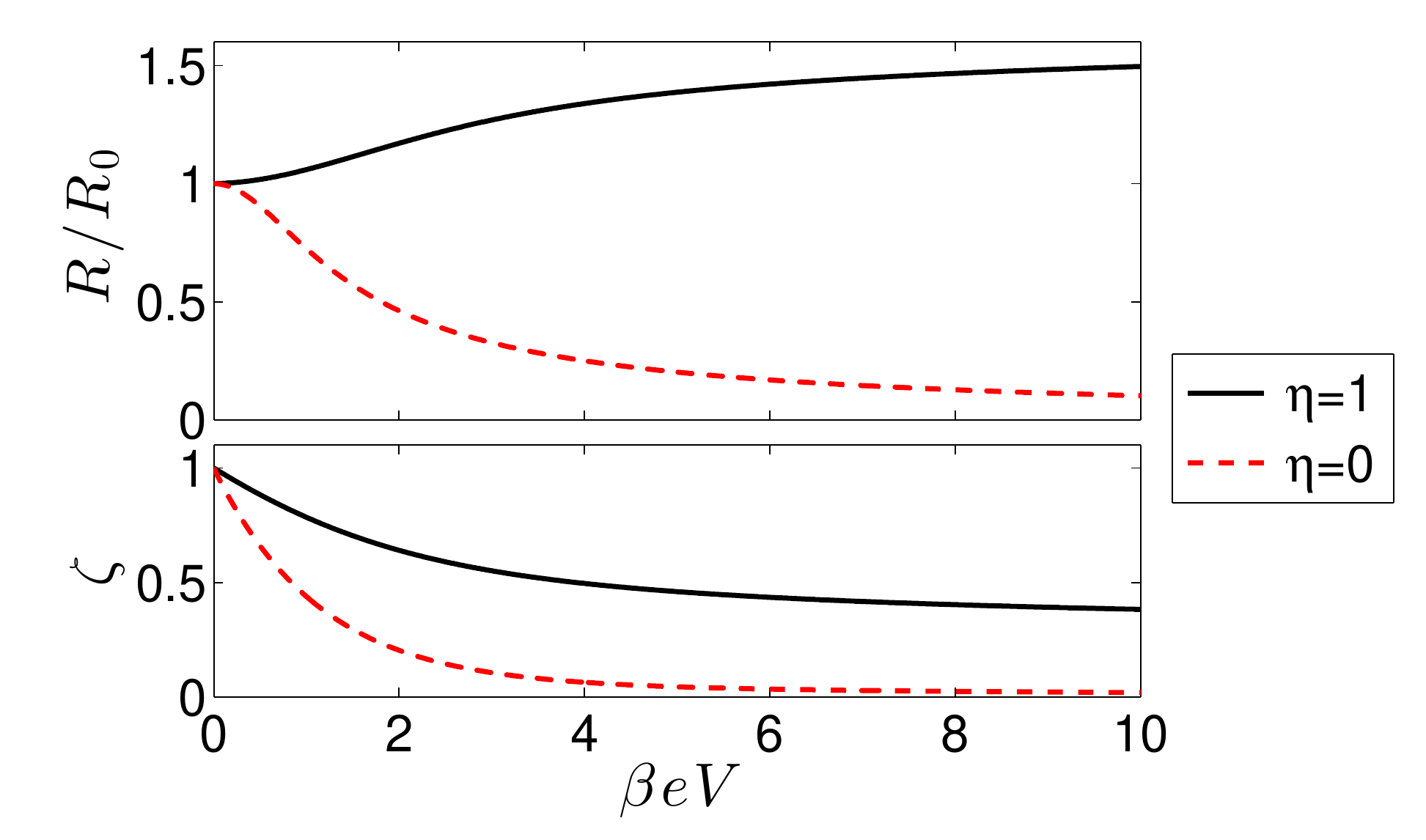}
  \caption{\label{fig:scenarios} The normalized impurity induced
    backscattering probability $R/R_0$ and $\zeta$, both in dependence
    of the ratio of transport voltage and temperature $\beta eV$.
    $R_0$ is the backscattering probability for $\beta eV \ll 1$.  In
    both cases, $S=5/2$, and
    $J_z^2/J_\perp^2 = J_\perp^2/J_\mathrm{aniso}^2 = 10$.  $R_0$ is
    enhanced by two orders of magnitude as the dimensionless strength
    of Rashba disorder $\eta$ is increased from $0$ to $1$.}
\end{figure}

\emph{Results:} Although the framework that we set up so far does not
rely on any specific assumptions about the couplings in \HS, it is
helpful to focus on the parameter regime
$J_z^2 \gg J_\perp^2 \gg J_\mathrm{aniso}^2$ for three reasons: (i)
From a microscopic analysis we found that this regime is
experimentally relevant for HgTe/CdTe quantum wells \cite{SuppMat}.
(ii) The importance of the Rashba disorder induced effective couplings
with regard to the dc resistance becomes particularly clear in this
parameter regime.  (iii) A clear hierarchy of couplings allows one to
disentangle the discussion of scattering processes.  The following
detailed discussion about the relevancy of Rashba disorder for the
impurity induced resistance leads to two important results: First,
while $R(\beta eV)$ is a monotonically decreasing function without
Rashba disorder, this monotonicity is reversed in the presence of
Rashba disorder (see Fig.~\ref{fig:scenarios}).  Second, $R_0$, the
backscattering probability in the limit $\beta eV \ll 1$, is
significantly increased by Rashba disorder.

Let us consider first the case where Rashba disorder is absent, i.e.,
$\eta=0$.  The only nonvanishing backscattering rate is $1/\tau_\perp$
from Eq.~\eqref{eq:tauperp}, which is small as it arises from an
interplay of scattering due to $J_\perp$ and $J_\mathrm{aniso}$.  In
particular, $R_0$ is found to be proportional to the harmonic mean of
$J_\perp^2$ and $2J_\mathrm{aniso}^2$, because
$p_\perp = [1+J_\perp^2/2J_\mathrm{aniso}^2]^{-1}$ for $\beta eV = 0$.
Consequently, when $J_\perp^2$ and $J_\mathrm{aniso}^2$ are very
different in magnitude, it is the smaller of the two which determines
the magnitude of $R_0$.  With increasing $\beta eV$, the
backscattering of right movers via $J_\perp S^+s^-$ becomes
increasingly dominant relative to other scattering rates and tends to
polarize the impurity, such that
$\zeta \simeq 2J_\mathrm{aniso}^2/\beta eV J_\perp^2$ approaches zero
in the large $\beta eV$ limit (cf. Fig.~\ref{fig:scenarios}).  With
increasing polarization, the probability for the impurity spin to be
in state $\ket{S}$ increases, and the probability for an individual
right-moving electron to be backscattered is suppressed.  This leads
to the monotonic decrease of $R(\beta eV)$ shown in
Fig.~\ref{fig:scenarios}, with $R \sim J_\mathrm{aniso}^2 / \beta eV$
in the large $\beta eV$ limit \cite{footnote:localMax}.

Rashba disorder, described by a finite $\eta$, has a profound effect
on the impurity induced resistance.  Since
$J_z^2 \gg J_\perp^2 \gg J_\mathrm{aniso}^2$, $1/\tau_{z,\mathrm{R}}$
dominates $1/\tau_\perp$ already for very weak Rashba disorder with
$\eta\gtrsim 4J_\mathrm{aniso}^2/J_z^2$.  The magnitude of $R_0$ is
then determined by $J_{z,\mathrm{R}}^2$ instead of
$J_\mathrm{aniso}^2$, which, depending on the precise values of the
couplings and $\eta$, can be a large difference.  Regarding the
dependence on the polarization, $1/\tau_{z,\mathrm{R}}$ is
qualitatively different from $1/\tau_\perp$.  Evaluating the sum over
$m$ in Eq.~\eqref{eq:tauzR} for the limiting cases of perfect
polarization, $P_m=\delta_{m,S}$, and vanishing polarization,
$P_m=1/(2S+1)$, yields $S^2$ in the former and $S(S+1)/3$ in the
latter case, respectively.  This shows that, for spin $S>1/2$, the
rate $1/\tau_{z,\mathrm{R}}$ increases with increasing polarization,
because $S^2 > S(S+1)/3$.  However, $1/\tau_{z,\mathrm{R}}$ is
independent of $\zeta$ for $S=1/2$.  Thus, in contrast to the case
without Rashba disorder, $R$ is now found to monotonically increase
with $\beta eV$ when $S>1/2$.

For real samples with several impurities, the total backscattering
rate is proportional to the number of impurities in the absence of
localization. Based on Eq.~\eqref{eq:resistance}, we estimate the mean
free path to be $4\,\mu$m in $7.0\,$nm wide HgTe quantum wells with a
lattice constant $a=0.65\,$nm and $v=4\times 10^5\,$m/s, by assuming
(i) that there is a concentration of $10^{-4}$ Mn$^{2+}$ ions
($S=5/2$) per unit cell, (ii) $\zeta=0.5$ and spatial average
$\langle J_z^2/\hbar^2v^2 \rangle = 0.035$ \cite{SuppMat}, and (iii)
$\eta \approx 3$ based on the estimate for $V_0$ from
Refs.~\cite{DelMaestro2013, Glazov2010}.

\begin{figure}
  \includegraphics[width=\columnwidth]{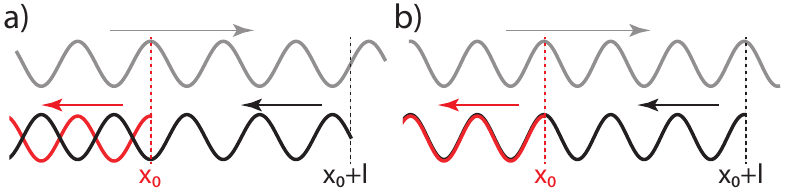}
  \caption{\label{fig:reflectedWaves} A right-moving wave (gray) is
    phase coherently backscattered from impurities at $x_0$ (red) and
    $x_0+l$ (black).  Destructive (constructive) interference occurs
    when $k_\mathrm{F}l=\nu\pi$ with half-integer (integer) $\nu$;
    cf. a) [b)].}
\end{figure}

A plane wave $e^{ik_\mathrm{F}x}$ with Fermi momentum $k_\mathrm{F}$
can be backscattered elastically from impurities at $x_0$ and $x_0+l$
by the process Eq.~\eqref{eq:MatrixElement} (see Fig.~3).  The partial
waves $e^{-ik_\mathrm{F}(x-x_0) +ik_\mathrm{F}x_0}$ and
$e^{-ik_\mathrm{F}(x-x_0-l)+ik_\mathrm{F}(x_0+l)}$ interfere,
resulting in a contribution to the backscattering probability
$ \propto \cos k_\mathrm{F}l$.  Tuning $k_\mathrm{F}$ via a back-gate
voltage alternately causes constructive or destructive
interference. Hence, conductance fluctuations occur, in agreement with
experiment.  Several impurities give rise to more complex
oscillations.

\emph{Summary:} We determined the dc resistance of helical edge states
in the presence of Rashba disorder and a magnetic impurity with spin
$S>1/2$.  As a key result, we find that combined scattering from both
Rashba disorder and impurity is enhanced as the impurity becomes more
polarized, giving rise to a resistance that slowly increases as the
ratio of transport voltage and temperature increases, in agreement
with experiments \cite{Koenig2007, Gusev2014, Yacoby:private}.  Since
backscattering is elastic, quantum interference can explain the
occurrence of conductance fluctuations.

\begin{acknowledgments}
  \emph{Acknowledgment:} We acknowledge valuable discussions with
  A.\,Yacoby, S.\,Hart, and B.\,I.\,Halperin, and financial support by
  ESF and DFG Grants No. RO 2247/7-1 and No. RO 2247/8-1.
\end{acknowledgments}

\onecolumngrid
\section*{Supplemental Material}
\subsection*{HgTe/CdTe quantum wells: microscopic considerations about
  \HS}
In many cases, it is a useful simplification to consider the edge
states of a topological insulator as one-dimensional objects by
neglecting the spatial extension of their wave function into the bulk.
Here, we take this spatial extension as well as the orbital character
of the edge states into account, in order to derive from the exchange
interaction of bulk electrons in HgTe or CdTe with a point-like
magnetic impurity, the effective exchange interaction between such an
impurity and electrons in the helical edge states of a topologically
non-trivial HgTe/CdTe quantum well.  In what follows, we choose
coordinates such that the $x$-axis coincides with the edge and the
$z$-axis is parallel to the growth direction of the quantum well.  The
location of the impurity is chosen to be $\vec{r}_0 = (0, y_0, z_0)$.

In Ref.~\cite{SM:Bernevig2006}, the BHZ Hamiltonian was derived, to
describe the low energy physics of HgTe/CdTe quantum wells.  This
effective Hamiltonian for the quantum well acts on the quantum well
subbands $\ket{E1,\pm}$ and $\ket{H1,\pm}$, which are the subbands
with the lowest energy at the $\Gamma$-point.  Hence, it is usually
written as a $4\times 4$ matrix in the basis
$(\ket{E1,+}, \ket{H1,+}, \ket{E1,-}, \ket{H1,-})$.  The BHZ
Hamiltonian was derived from the six-band bulk Kane model, which is
formulated in the basis
$(\ket{\Gamma_6, +1/2}, \ket{\Gamma_6, -1/2}, \ket{\Gamma_8, +3/2},
\ket{\Gamma_8, +1/2}, \ket{\Gamma_8, -1/2}, \ket{\Gamma_8, -3/2})$.
As a consequence $\ket{E1,\pm}$ and $\ket{H1,\pm}$ can each be
represented in the basis of the six-band Kane model:
$\ket{E1,+} = f_1(z)\ket{\Gamma_6,+1/2} + f_4(z)\ket{\Gamma_8,+1/2}$,
$\ket{H1,+} = f_3(z)\ket{\Gamma_8,+3/2}$ \cite{SM:Bernevig2006}.  Edge
states, which can be found as eigenstates of the BHZ Hamiltonian with
appropriate boundary conditions \cite{SM:Zhou2008, SM:Wada2011}, thus
can be expressed in the basis of the six-band Kane model.  To be
precise, we express the edge state $\ket{k,\sigma}$ in position space
\begin{equation}
  \langle\vec{r} \ket{k,\sigma} = \frac{1}{\sqrt{L}} e^{ikx}
  M_\mathrm{bulk}(z) M_\mathrm{qw}(k,y) \ket{\sigma},
\end{equation}
where the spinors $\ket{\up} = (1,0)^T$ and $\ket{\down} = (0,1)^T$
correspond to right- and left-movers, respectively.  The matrices
$M_\mathrm{bulk}$ and $M_\mathrm{qw}$ explicitly read
\begin{equation}
  M_\mathrm{bulk}(z) \equiv
  \begin{pmatrix}
    f_1(z) & 0 & 0 & 0\\
    0 & 0 & -f_1^*(z) & 0\\
    0 & f_3(z) & 0 & 0\\
    f_4(z) & 0 & 0 & 0\\
    0 & 0 & f_4^*(z) & 0\\
    0 & 0 & 0 & -f_3^*(z)
  \end{pmatrix},\qquad
  M_\mathrm{qw}(k,y) \equiv
  g_k(y)
  \begin{pmatrix}
    c_{E1} & 0 \\
    c_{H1} & 0 \\
    0 & c_{E1}^* \\
    0 & c_{H1}^*
  \end{pmatrix}.
\end{equation}
The $4\times 2$ matrix $M_\mathrm{qw}$ is used to express the spinors
$\ket{\sigma}$ in the basis of the BHZ Hamiltonian, while the
$6\times 4$ matrix $M_\mathrm{bulk}$ expresses the elements of this
basis in the basis of the six-band Kane model.  The components
$c_{E1}$ and $c_{H1}$ can be obtained analytically \cite{SM:Zhou2008,
  SM:Virtanen2012}, we use $c_{E1}^2 = 0.13$ and $c_{H1}^2 = 0.87$
independent of $k$.  In addition, $g_k(y)$ describes the exponential
decay of the edge state wave function away from the edge; it is known
analytically \cite{SM:Zhou2008, SM:Wada2011}.  The functions $f_i(z)$
were described in Ref.~\cite{SM:Bernevig2006}.  $f_1$ and $f_3$ are
even and real, whereas $f_4$ is odd and imaginary; all three functions
decay exponentially in the CdTe region.

The interaction of bulk electrons in CdTe and HgTe with the local
moment of a magnetic impurity like a Mn$^{2+}$ ion may be described by
the Hamiltonian
$H_\mathrm{ex}^\mathrm{bulk} = \alpha \vec{S}\cdot\vec{s}_{\Gamma_6} +
(\beta/3) \vec{S}\cdot \vec{I}_{\Gamma_8}$,
\cite{SM:Merkulov2010}.  Here, $\vec{S}$ acts on the local moment,
$\vec{s}_{\Gamma_6}$ is the local spin density of electrons in the
$\Gamma_6$ band at the impurity site, $\vec{I}_{\Gamma_8}$ is the
density of total angular momentum of holes in the $\Gamma_8$ band
evaluated at the impurity site, and $\alpha$ and $\beta$ are exchange
constants.  Expressing $H_\mathrm{ex}^\mathrm{bulk}$ in the basis of
the six-band Kane model,
$\ket{\Gamma_6, +1/2}, \ket{\Gamma_6, -1/2}, \ket{\Gamma_8, +3/2},
\ket{\Gamma_8, +1/2}, \ket{\Gamma_8, -1/2}, \ket{\Gamma_8, -3/2}$,
and keeping in mind that the  operators $S^+$,
$S^-$, and $S^z$ act on the impurity state, we find
\begin{equation}
  H_\mathrm{ex}^\mathrm{bulk} =
  \left(
    \begin{array}{c|c}
      \alpha\frac{1}{2}\begin{pmatrix}
        S^z & S^- \\
        S^+ & -S^z
      \end{pmatrix}
      &  \\\hline
      &
        \frac{\beta}{3} \frac{1}{2}
        \begin{pmatrix}
          3S^z & \sqrt{3}S^-  & 0 & 0 \\
          \sqrt{3}S^+ & S^z & 2S^- & 0\\
          0 & 2S^+ & -S^z & \sqrt{3}S^- \\
          0 & 0 &\sqrt{3}S^+ & -3S^z\\
        \end{pmatrix}
    \end{array}
  \right)
  \delta(x) \delta(y-y_0) \delta(z-z_0).
\end{equation}

We are interested in the effective exchange interaction
$H_\mathrm{ex}^\mathrm{edge}$ between the edge electrons and the
impurity.  To obtain the corresponding $2\times 2$ Hamiltonian in the
basis $(\ket{\up}, \ket{\down})$, we first calculate
$H_\mathrm{ex}^\mathrm{qw}(x,y) = \int dz\, M_\mathrm{bulk}^\dagger
H_\mathrm{ex}^\mathrm{bulk} M_\mathrm{bulk}$
to find the effective exchange interaction in the basis of the BHZ
Hamiltonian and afterwards compute
$H_\mathrm{ex}^\mathrm{edge}(x) = \int dy\, M_\mathrm{qw}^\dagger
H_\mathrm{ex}^\mathrm{qw} M_\mathrm{qw}$.
\begin{equation}
  H_\mathrm{ex}^\mathrm{qw}(x,y) =
  \int dz\, M_\mathrm{bulk}^\dagger H_\mathrm{ex}^\mathrm{bulk} M_\mathrm{bulk} =
  \begin{pmatrix}
    J^\mathrm{qw}_1 S^z & (J^\mathrm{qw}_4)^* S^+  & (J^\mathrm{qw}_3)^* S^- & 0 \\
    J^\mathrm{qw}_4 S^- & J^\mathrm{qw}_2 S^z & 0 & 0 \\
    J^\mathrm{qw}_3 S^+ & 0 & -J^\mathrm{qw}_1 S^z & -J^\mathrm{qw}_4 S^- \\
    0 & 0 & -(J^\mathrm{qw}_4)^* S^+ & -J^\mathrm{qw}_2 S^z
  \end{pmatrix}
  \delta(x)\delta(y-y_0),
\end{equation}
with
\begin{subequations}
  \begin{align}
    J^\mathrm{qw}_1(z_0) &= \frac{1}{6}( 3\alpha|f_1(z_0)|^2 + \beta|f_4(z_0)|^2), \\
    J^\mathrm{qw}_2(z_0) &= \frac{1}{2} \beta |f_3(z_0)|^2, \\
    J^\mathrm{qw}_3(z_0) &= \frac{1}{6}( -3\alpha f_1(z_0)^2 + 2\beta f_4(z_0)^2), \\
    J^\mathrm{qw}_4(z_0) &= \frac{1}{2\sqrt{3}} \beta f_3^*(z_0) f_4(z_0).
  \end{align}
\end{subequations}
\begin{equation}
  H_\mathrm{ex}^\mathrm{edge}(x) =
  \int dy\, M_\mathrm{qw}^\dagger H_\mathrm{ex}^\mathrm{qw} M_\mathrm{qw} =
  \begin{pmatrix}
    J_z S^z + J_\mathrm{aniso} S^- + J_\mathrm{aniso}^* S^+ & J_\perp^* S^- \\
    J_\perp S^+ & -(J_z S^z + J_\mathrm{aniso} S^- + J_\mathrm{aniso}^* S^+)
  \end{pmatrix}
  \delta(x)
\end{equation}
with
\begin{subequations}
  \begin{align}
    J_z(y_0, z_0) &= |g_k(y_0)|^2 \left(J^\mathrm{qw}_1 |c_{E1}|^2 + J^\mathrm{qw}_2 |c_{H1}|^2\right) \\
    J_\perp(y_0, z_0) &= |g_k(y_0)|^2 J^\mathrm{qw}_3 c_{E1}^2 \\
    J_\mathrm{aniso}(y_0, z_0) &= |g_k(y_0)|^2 J^\mathrm{qw}_4 c_{E1}c_{H1}^*
  \end{align}
\end{subequations}

$H_\mathrm{ex}^\mathrm{edge}$ is equivalent to \HS\ defined in Eq.~(3)
of the main text, except that the complex phase of $J_\mathrm{aniso}$
was neglected in the main text for brevity; $J_\perp$ is real.  Values
for $\alpha$ and $\beta$ can be found in the literature for HgTe as
well as CdTe \cite{SM:Merkulov2010, Furdyna1988}.  $J_z$ is found to
be the largest of the three couplings independent of $z_0$.  $f_3(z)$
is odd such that $J_\mathrm{aniso}$ vanishes if the impurity is
located at $z_0=0$, i.e. in the middle of the quantum well.
Consequently $J_z^2 \gg J_\perp^2 \gg J_\mathrm{aniso}^2$ is realistic
for $z_0\approx 0$.

The reason why a finite $J_\mathrm{aniso}$ can arise from an isotropic
exchange interaction like $H_\mathrm{ex}^\mathrm{bulk}$ can be
understood from the calculation and was explicitly stated in
Ref.~\cite{SM:Maciejko2011}: The electrons in the edge states in
HgTe/CdTe quantum wells are \emph{not} spin-polarized because of the
spin-orbit interaction in these materials. This also holds for
InAs/GaSb/AlSb quantum wells.

\subsection*{Details and intermediate steps regarding the effective
  coupling $J_{z,\mathrm{R}}$}
In this section we provide the details of the determination of the
effective coupling $J_{z,\mathrm{R}}$ in Eq.~(5) of the main text, by
calculating the disorder averaged squared modulus of matrix elements
of the $T$-matrix term
$T_{z,\mathrm{R}}(\epsilon) = \HR G(\epsilon) J_zS^zs^z + J_zS^zs^z
G(\epsilon) \HR$
for elastic backscattering.  Let us calculate explicitly the matrix
element of the first term, which is represented in Fig.~1\,a) of the
main text, the contribution from the second term is obtained
analogously.  We use the Fourier representations
$J_z S^zs^z = \frac{J_z}{L}S^z \sum_{k_1,k_2} c_{k_1,\alpha}^\dagger
\sigma_{\alpha\alpha}^z c_{k_2,\alpha} $
and
$\HR = \frac{-1}{\sqrt{L}} \sum_{k_1,k_2} (k_1+k_2)a_{k_1-k_2}
c_{k_1,\alpha}^\dagger \sigma^y_{\alpha\beta} c_{k_2,\beta}$ to find
\begin{equation}
\begin{split}
  &\!\!\!
  \bra{m;-k,-\sigma} \HR \frac{1}{\sigma v\hbar k - H_0} J_zS^zs^z\ket{m;k,\sigma} \\
  &= \sum_{k'}
  \frac{\langle m;-k,-\sigma| \frac{-\sigma^y_{\alpha\beta}}{\sqrt{L}} \sum_{k_1,k_2}
    (k_1+k_2)a_{k_1-k_2} c_{k_1,\alpha}^\dagger c_{k_2,\beta}
     | m;k',\sigma \rangle
    \langle m;k',\sigma| \frac{J_z}{L}S^z\sigma_{\alpha\alpha}^z \sum_{k_1,k_2}
    c_{k_1,\alpha}^\dagger c_{k_2,\alpha} |
    m;k,\sigma \rangle}{\sigma\hbar v(k-k')} \\
  &= \sum_{k'}
  \frac{\langle m;-k,-\sigma| \frac{-\sigma^y_{-\sigma\sigma}}{\sqrt{L}}
    (-k+k')a_{-k-k'} c_{-k,-\sigma}^\dagger c_{k',\sigma}
     | m;k',\sigma \rangle
    \langle m;k',\sigma| \frac{J_z}{L}S^z\sigma_{\sigma\sigma}^z
    c_{k',\sigma}^\dagger c_{k_k,\sigma} |
    m;k,\sigma \rangle}{\sigma\hbar v(k-k')} \\
  &= \frac{\sigma i J_z}{L\sqrt{L}\hbar v} \bra{m}S^z \ket{m}
  \sum_{k'} a_{-k-k'}\quad.
  \end{split}
\end{equation}
The second term yields almost the same result, except that the sum now
is $\sum_{k'} a_{k'-k}$.  Hence for the disorder averaged squared
modulus of the matrix element, which enters into the scattering rates,
we find
\begin{equation}
\begin{split}
  \langle |\bra{m;-k,-\sigma}T_{z,\mathrm{R}}\ket{m;k,\sigma}|^2 \rangle_\mathrm{dis} &= 
  \frac{1}{L^3} \frac{J_z^2}{\hbar^2 v^2}  |\bra{m}S^z \ket{m}|^2
  \langle \sum_{k'} (a_{k'-k} + a_{-k-k'}) \sum_{k'}(a_{k'-k} + a_{-k-k'})^*\rangle_\mathrm{dis} \\
  &= \frac{1}{L^2} 4\eta J_z^2  |\bra{m}S^z \ket{m}|^2
\end{split}
\end{equation}
where $\eta = V_0 / \hbar^2 v^2$.  We used
$\langle a_q a_{q'}^*\rangle_\mathrm{dis} =
V_0\delta_{q,q'}\tilde{F}(q)$,
with $\tilde{F}(q) = \frac{1}{L}\int_0^L \mathrm{d}x\, F(x)e^{-iqx}$
and $F(0)=1$.  Since
$|\bra{m;-k,-\sigma} J_{z,\mathrm{R}} S^z(s^++s^-) \ket{m; k,
  \sigma}|^2 = \frac{1}{L^{2}} J_{z,\mathrm{R}}^2 |\bra{m}S^z
\ket{m}|^2$,
we can capture the effects from the second order process on the
scattering rates by considering first order processes from the
effective coupling Eq.~(5).

\subsection*{Explicit expressions for $\zeta$ and $p_\perp$.}
As mentioned in the main text $\zeta = \Gamma_{m-1,m}/\Gamma_{m,m-1}$.
Since $|\bra{m-1}S^-\ket{m}|^2 = |\bra{m}S^+\ket{m-1}|^2$, these
matrix elements cancel.  Consequently $\zeta$ does not depend on $m$,
but only on the (effective) coupling constants and on $\beta eV$
through the integrals $I^\sigma$
\begin{equation}
  \zeta =  \frac{
    2(J_\mathrm{aniso}^2 + J_{\perp,\mathrm{R}}^2) + I^\down(J_\perp^2 + J_{\mathrm{aniso},\mathrm{R}}^2) + I^\up J_{\mathrm{aniso},\mathrm{R}}^2
  }{
    2(J_\mathrm{aniso}^2 + J_{\perp,\mathrm{R}}^2) + I^\down J_{\mathrm{aniso},\mathrm{R}}^2 + I^\up (J_\perp^2 + J_{\mathrm{aniso},\mathrm{R}}^2)
  }.
\end{equation}

The probability $p_\perp$ is calculated according to the following
considerations: Assume a right-moving electron is backscattered while
flipping the impurity from $m$ to $m+1$, i.e. assume that a
backscattering event of the type
$\ket{m;\up}\ \rightarrow\ \ket{m + 1;\down}$ occurred.  Then, because
the impurity is assumed to be in a stationary state, at some time
another scattering event
$\ket{m+1;\sigma'}\ \rightarrow\ \ket{m;\sigma}$ has to take place.
If the second event happens to be a backscattering of a left-moving
electron ($\sigma'=\down$, and $\sigma=\up$), then in total, the two
scattering events left the current transmitted between the two
reservoirs unchanged.  However, if the second event is a forward
scattering event ($\sigma'=\sigma$) or the backscattering of another
right-mover ($\sigma'=\up$, and $\sigma=\down$), then the first
backscattering event indeed had a net effect on the current and thus
contributed to the dc resistance.  Since it does not matter which
momentum the states involved in the scattering have, the probabilities
for the two different cases are proportional to the integrated
scattering rates $\Gamma_{mm'}^{\sigma\sigma'}$.  The probability that
the first scattering event \emph{does not} have a net effect on the
resistance is
$p^\mathrm{neg}_{m+1,m} \propto \Gamma_{m,m+1}^{\up\down}$.  On the
other hand, the probability that the first scattering event
\emph{does} have a net effect on the resistance is
$p^\mathrm{pos}_{m+1,m} \propto \Gamma_{m,m+1}^{\up\up} +
\Gamma_{m,m+1}^{\down\down} + \Gamma_{m,m+1}^{\down\up}$.
Since $p^\mathrm{pos}_{m+1,m} + p^\mathrm{neg}_{m+1,m} = 1$ we find
\begin{equation}
  p^\mathrm{pos}_{m+1,m} = \left( 1 + \frac{\Gamma_{m,m+1}^{\up\down}}{\Gamma_{m,m+1}^{\up\up} +
      \Gamma_{m,m+1}^{\down\down} + \Gamma_{m,m+1}^{\down\up}} \right)^{-1}.
\end{equation}
Since $\Gamma_{m,m+1}^{\sigma\sigma'} \propto |\bra{m}S^-\ket{m+1}|^2$
independently of $\sigma$ and $\sigma'$, $p^\mathrm{pos}_{m+1,m}$ is
independent of $m$ and we can define
\begin{equation}
p_\perp\equiv p^\mathrm{pos}_{\frac{1}{2},-\frac{1}{2}} =  \left( 1 +
    \frac{(J_\perp^2 + J_{\mathrm{aniso},\mathrm{R}}^2) I^\down}{2(J_\mathrm{aniso}^2 + J_{\perp,\mathrm{R}}^2) +
                  J_{\mathrm{aniso},\mathrm{R}}^2I^\up}
    \right)^{-1}.
\end{equation}


\begin{thebibliography}{99}
\bibitem{footnote:tau2} In Eq.~\eqref{eq:resistance}, we omit a third
  scattering rate arising from the effective coupling
  $J_{\mathrm{aniso},\mathrm{R}} (S^+s^+ + S^-s^-)$, because its
  impact on the backscattering probability is negligible in all
  situations that we are interested in.
\bibitem{footnote:localMax} We suppressed in our discussion the fact
  that $p_\perp$ increases with $\beta eV$ and approaches $1$ for
  $\beta eV\gg 1$.  However, a detailed analysis shows that
  $R(\beta eV)$ does not have a local maximum, i.e., effects due to
  varying polarization dominate.
\end{thebibliography}
\end{document}